\newcommand{\nii}{[N\,{\sc ii}]}

\newcommand{\ha}{${\rm H \alpha}$}

\documentclass[]{spie}  
\usepackage[]{graphicx}
\usepackage{url}

\title{Small Solutions to the Large Telescope Problem: A Massively 
Replicated MEMS Spectrograph} 

\author{Nicholas P. Konidaris II\supit{a}, Joel A. Kubby\supit{b} and 
Andrew I. Sheinis\supit{c}
\skiplinehalf
\supit{a} Department of Astronomy and Astrophyiscs, UCSC, Santa Cruz, 
CA; \\
\supit{b} Department of Electrical Engineering, UCSC, Santa Cruz, CA; \\
\supit{c} Astronomy Department, University of Wisconsin, Madison, WI
}

\begin{document} 
  \maketitle

\begin{abstract}

In traditional seeing-limited observations the spectrograph aperture 
scales with telescope aperture, driving sizes and costs to enormous 
proportions. We propose a new solution to the seeing-limited spectrograph problem. A massively fiber-sliced configuration feeds a set of small diffraction-limited spectrographs. We present a prototype, tunable, J-band, diffraction grating, designed specifically for Astronomical applications: The grating sits at the heart of a spectrograph, no bigger than a few inches on a side. Throughput requirements dictate using tens-of-thousands of spectrographs on a single 10 to 30 meter telescope. A full system would cost significantly less than typical instruments on 10m or 30m telescopes.

\end{abstract}

\keywords{astronomy, MEMS, seeing limited, fiber slicing, NIR, 
tunable diffraction grating}

\section{INTRODUCTION}
\label{sec:intro}
The next generation of large telescopes is designed, among other scientific requirements, to measure fundamental properties of galaxies from high-redshift ($z$) to today.  However, the instrumentation cost for the next generation of telescopes will be enormous.  Here, we present an approach that combines several modern technologies to significantly reduce the cost of astronomical spectrographs.

\begin{figure}[htp]
\includegraphics{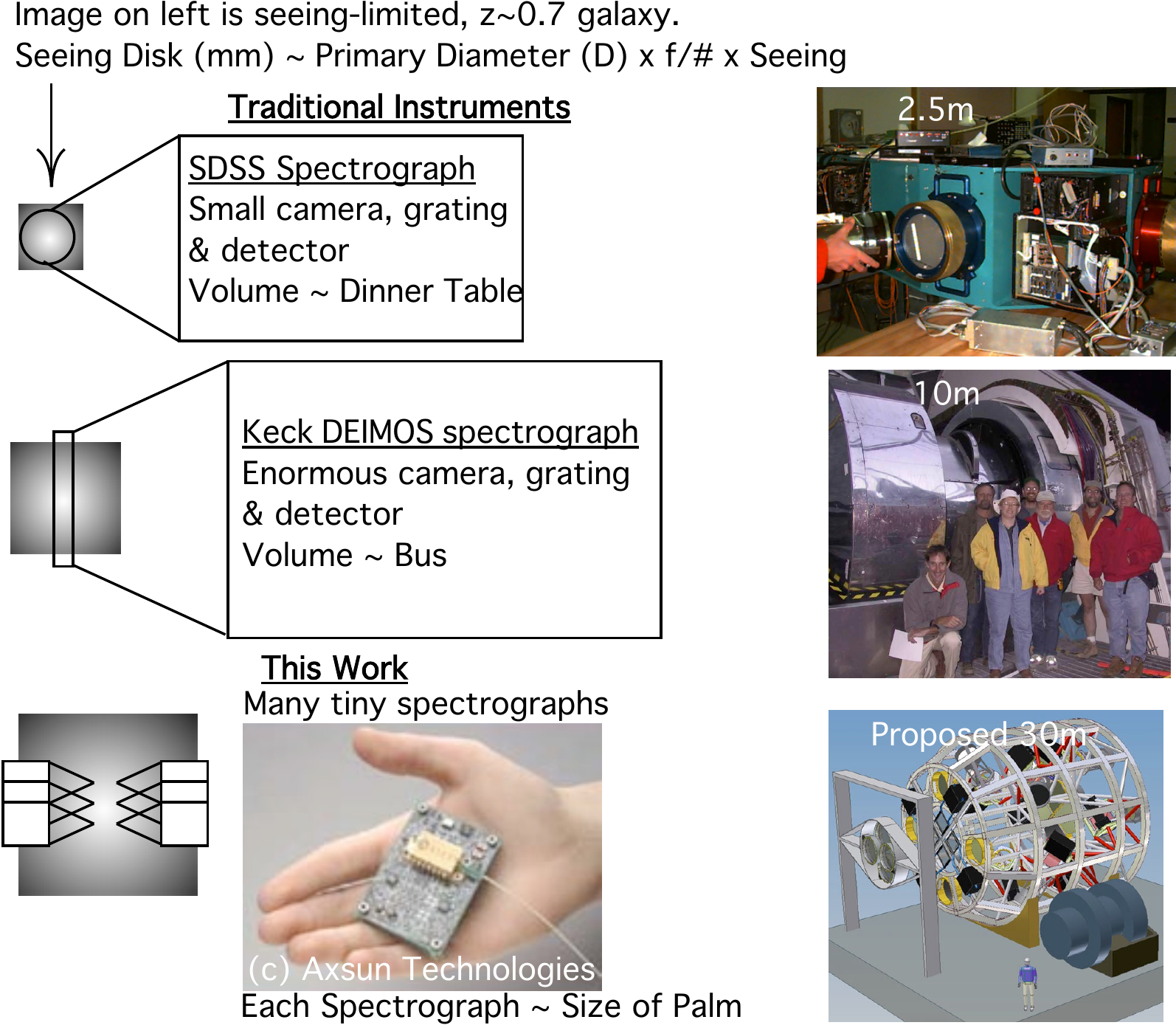}
\caption{\label{fig:cost}A schematic illustration of the evolution in size and cost of optical spectrgraphs.  The top two rows show pictures of the Sloan digital Sky Survey  (SDSS; 2.5m primary) multi-object  spectrograph, and the Keck (10m primary) spectrograph, DEep Imaging Multi-Object Spectrograph.  Notice the relationship between telescope size (going down the page) and instrument volume.  At 10m, the spectrograph dwarfs the PI and engineers.  For the proposed 30m telescope and wide field optical spectrograph (WFOS), a CAD model is rendered next to DEIMOS.  The WFOS is the size of a small house. The device in the bottom left is a Fabry-Perot spectrograph built by Axsun Technologies, it illustrates the principle behind this design: build thousands of miniature spectrographs.}
\end{figure}

For any two functionally equivalent spectrographs, their volume increases, roughly, with the telescope diameter cubed\cite{Schroeder87}.   There is a power-law relationship between telescope diameter (volume) and instrument cost, where the exponent sits between two and three. For example, DEIMOS, on a 10m telescope cost US\$10m dollars\cite{faber03} (though it was originally budgeted for less.)  The preliminary estimate for an instrument with less multiplexing capability, on a 30m telescope, is US\$60m\cite{paz06}.  This growth of telescope diameter, and instrument cost and complexity, is schematically illustrated in Figure \ref{fig:cost}. 

In this paper we describe the design of a spectrograph, built with tens of thousands of mini-spectrographs.  By clever use of fiber multiplexing and slicing, enormous optics disappear, simplifying alignment and test procedures, which ultimately drive down the cost of the instrument package.  A single, MEMS, spectrograph, including detector, costs manufacturers roughly US\$20 per part.

It is well understood that industrial replication reduces the cost of an individual component.  However, a surprising recent result, is that replicated spectrographs reduce the {\em total} cost of an instrument.  For example, the recently designed Visible Integral-Field Replicable Unit Spectrographs (VIRUS), which is replicated by $\sim100\times$, saves approximately a factor of two or three compared to a monolithic design\cite{hill06}.

Micro-electrical-mechanical-systems (MEMS) technology, allows the production of spectrographs even smaller than the VIRUS.  The drive to reduce spectrograph size is the drive reduce total instrument cost.  Though these MEMS spectrographs may sound exotic, we emphasize that this technology is not new\cite{onat05, Musca05}, commercial, off-the shelf MEMS spectrographs exist today.  However, these commercial spectrographs, are typically tuned for specific applications (e.g. laser communication, television displays and molecular identification.)   This work leverages the research of MEMS spectroscopy, to design an instrument tuned for a specific astronomical application.

\section{SYSTEM DESIGN}

Our science goal is best accomplished by measuring the strength of the emission lines \ha\ (6563\AA) and \nii\ at $z\sim1$.  At this redshift, the emission lines fall into the NIR ($1.3\mu$.)  In the NIR, night-sky emission lines dominate the sky-background, so the spectrograph must have sufficient resolution to split night-sky lines, and resolve emission lines in a 100 ${\rm km s^{-1}}$ galaxy -- dictating a spectral resolution $R\sim4000$.

Further, there is a trade-off between detector real-estate (in terms of pixels, and dollars), and science.  To maximize science and minimize cost, we focus on the emission lines \ha\ and \nii.  At the redshift of interest, the total rest wavelength coverage necessary is 100\AA.  Linear InGaAs arrays\cite{nel06} with 128 pixels will satisfy the wavelength coverage and resolution requirements.

With only 100\AA\ of coverage, the diffraction grating's blaze function must be tunable.  The wavelength coverage of an InGaAs NIR linear area extends from $1-1.5\mu$.  Ultimately, a grating capable of tuning over this whole range would be valuable, however, as proof-of-concept, the ability to tune over the range of maximum sensitivity of the dector, $1.2-1.4\mu$, is sufficient.

Given our above requirements, we describe the designed system.  Starting at the focal plane of the telescope, we cover a single target with a fiber bundle in an integral-field configuration, each fiber is sized to the diffraction limit of the telescope\cite{Sheinis06}.  For each fiber, there is a single MEMS spectrograph.  So, for any one object in the focal plane, there may be 100 spectrographs placed on it.  As many integral-field bundles as packaging and positioning allow can be used, though, cost savings through economies of scale dictate that there should be at least $\sim100$ such integral field units.\footnote{We parenthetically emphasize that prototyping from the ground-up is the natural with multiplexed spectrographs, and stands in stark contrast to one-off monolithic designs.}  Bundles are placed with some kind of fiber-positioning robot.  

During observation a target will be assigned an integral-field bundle of $\sim100$ spectrographs.  The blaze-function of their gratings is tuned to peak at the observed wavelength of \ha\ ($\lambda_{obs}=(1+z)\cdot6563$\AA.)   As a result, a photometric- or spectroscopic- $z$ must be known for each target.  Typically, for NIR observations, $z$ is predetermined, to ensure that lines of interest do not fall on night-sky lines.

After this brief review of requirements and overall system design, we focus our attention on the individual, MEMS, spectrographs.  Each spectrograph is a scaled-down version of a typical grating spectrograph.  The light path begins at the fiber feed, the beam is collimated, then diffracted off of a tunable, MEMS, diffraction, grating, and finally, is brought to focus on a linear InGaAs array\cite{nel06}.  This research has focused on designing, simulating and building a tunable diffraction grating.  Examples of the miniaturized optics and assembly procedure has been discussed in \S1.  The innovation of this work is in producing a tunable grating that satisfies these above requirements.

\section{THE TUNABLE DIFFRACTION GRATING}
\label{S:grating}
There are a number of ways to manipulate a grating to manipulate its blaze function.  The grating equation relates the output angle, $\theta_{out}$, of a wavelength of light:
\begin{equation}
\sin\left(\theta_{out}\right)=\frac{m\cdot\lambda}{d}-\sin\left(\theta_{in}\right)
\end{equation}
to a variety of tunable parameters: $\theta_{in}$, the input angle of the beam, $\lambda$ is the beam's wavelength, $m$ is the operating order, and $d$ is the grating spacing.   Of these tunable parameters,  $\theta_{out}$ is most sensitive to changes in $d$:
\begin{equation}
\frac{\partial\theta_{out}}{\partial d} \propto d^{-2}
\end{equation}
As a result, our MEMS grating adjusts $d$ to tune the output angle.

To build our device, we fabricated our design on a recent PolyMUMPS\texttrademark\ run. PolyMUMPS is a commercial, low-cost, MEMS fabrication process, which is ideal for proof-of-concept prototype designs.  Devices produced via PolyMUMPS comprise of two released layers of polysilicon (either 2 or $1.5\mu m$ in thickness), and one thin ($0.5\mu m$) gold layer.  In addition, a variety of sacrificial layers are used for forming structures\cite{car05}.

\begin{figure}[htp]
\begin{centering}
\includegraphics[width=3in]{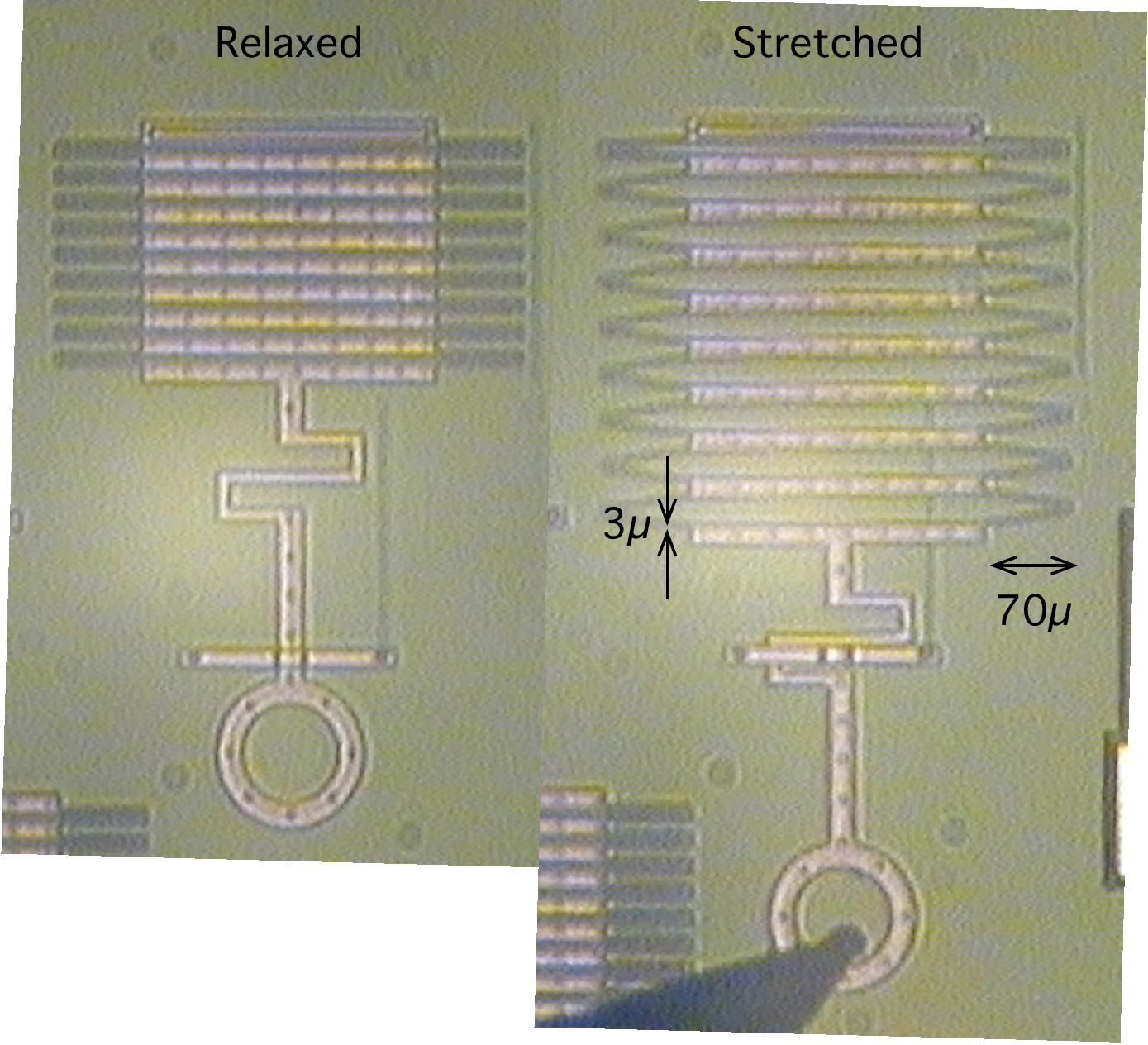}
\caption{\label{fig:grating}A small grating test-structure.  On the left is the relaxed grating, anchored to the substrate by the anchor blade on top, the remainder of the grating is floating.  On the right is the same grating, stretched by an off-chip probe (black shadow.)  This figure illustrates the basic principle of the diffraction grating ``accordion'' operation.  Thick blades are the diffracting elements, while thin springs ($3\mu$) thick connect the thick blades. Not shown in this diagram is the diffraction grating connection to the drive system by replacing the probe hole with the drive shuttle.}
\end{centering}
\end{figure}

To tune the grating spacing, the grating is designed to stretch in an ``accordion'' style.  Figure \ref{fig:grating} shows a test grating operated by hand.  This Figure is taken through a microscope objective.  A grating is comprised of thick diffraction blades, blades are made of polysilicon (Poly1), the total thickness of a blade is $2\mu m$.  Blades are connected via long, skinny springs, each $3 \mu m$ thick.  In this figure, the grating is stretched, ``by hand.''  That is, an off-chip probe, driven by hand turned screws pulls on the large circle.

We explored a number of spring designs by varying spring sizes.  Using the Intellisuite(tm) finite element analysis (FEA) code, we simulated displacements and measured spring constants, as well as maximum stress on the polysilicon, in steady state.  Typical simulated spring constants varied from $7.3 N\,m^{-1}$ to $61.0 N\,m^{-1}$. 

Aside from the spring constant, the design attempts to minimize the stress per unit displacement.  Our grating is fully compliant, so at sharp edges, the polysilicon fracture strength of $\sim1.5GPa$ must be considered.  Though smaller features increase fracture strength, the PolyMUMPS design rules dictate a minimize feature size of $2\mu$ and recommend at least $3\mu$ features to ensure that the lithographic process does not affect the features.  As a result, our final design has springs that are $3\times70 \mu m$ (as marked on Figure \ref{fig:grating}.)

As part of our inspection process, we attempted to measure the spring constant of our design.  An inhomogenous spring constant would result because polysilicon deposition may have varying material properties  over the wafer.  Our measurements were performed using the lab microscope, but were not able to detect any differences, largely due to the large, $2\mu m$, uncertainty in measurements.  Future work will use an interferometric approach.

Our diffraction grating is connected to a drive system described in the next section.  To envision this linkage, replace the probe hole seen in Figure \ref{fig:grating} with a connection to the actuation system. 

\subsection{Grating Diffraction Efficiency}
We use gsolver to simulate the efficiency of an alternating gold, polysilicon, diffraction grating.  The simulation takes account of the etch release holes, and dimples, visible in Figure \ref{fig:grating}.  Simulations showed acceptable efficiency in first order, typically greater than $50\%$.

Most of our efficiency loss is due to the design rules dictated by the PolyMUMPS process.  PolyMUMPS is a robust, and relatively simple service, designed to be general purpose with high yield.  As a result, our gratings are limited to square blaze functions.  Square blaze angles reduce the efficiency of the grating.  In \S\ref{S:risk} we briefly describe an alternative approach that increases simulated efficiency to greater than $85\%$.

\section{THE DRIVE SYSTEM} 

The drive system must deal with the trade off between actuator force, and total displacement.  Recall from the prior section, the displacement is directly responsible for the tunability of our system.  The tunable diffraction grating requires a large displacement of $\sim100\mu m$, under strong force, $\sim 10\mu N$.  MEMS actuators trade-off either force, or displacement.  As a result, we combine a set of small-stroke, yet high-force, actuators, and a ratcheting system\cite{Kolesar04}.

Two classes of thermal actuators are used in our design.  Thermal actuators deflect during thermal expansion induced by Joule heating.  To capitalize on deflection, asymmetries in the mechanical design yield directional net forces.  Because of small thermal masses, temperature changes are rapid, and these devices can be controlled accurately with electric currents.

\begin{figure}[htp]
\begin{centering}
\includegraphics{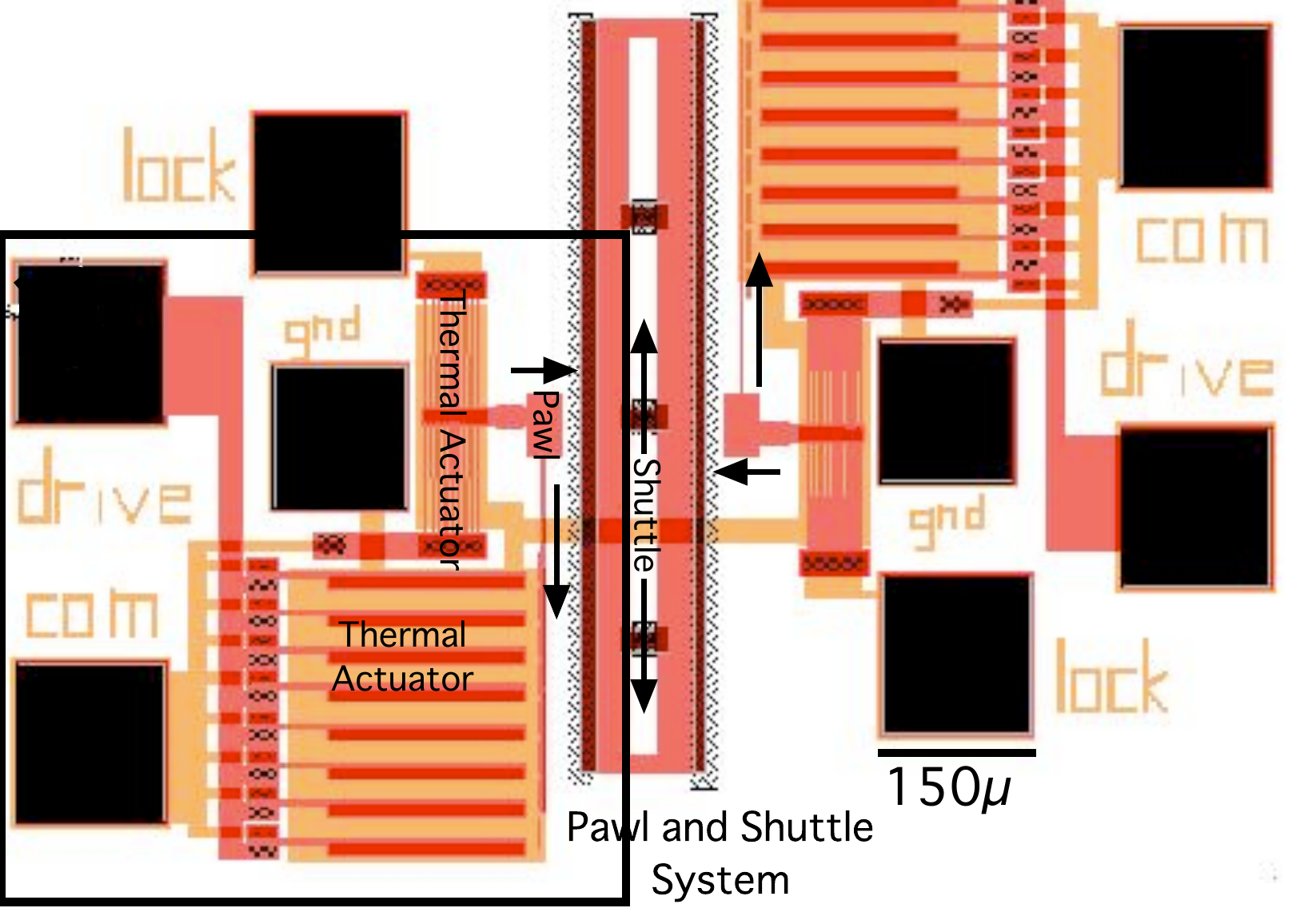}
\caption{\label{fig:device-drawing} Tanner EDA/MEMS design layout with major components labeled in black.  Starting from center, and working out clockwise within the large black box, we see:  The shuttle system, which, is connected to, and stretches the diffraction grating.  Left of the shuttle is a pair of thermal actuators.  Thermal actuators only actuate in one direction, as indicated by arrows.  Each side drives the shuttle in one direction only, and provides a locking mechanism.  Finally is the pawl, this device grips the shuttle through gears.  Figure \ref{fig:zoom1} zooms in to the black box.}
\end{centering}
\end{figure}

This system design is shown in Figure \ref{fig:device-drawing}.  This image is taken directly from the Tanner\texttrademark/EDA layout package.  In the Tanner/EDA design drawings, the first polysilicon layer (Poly0) is shown in orange, the third polysilicon layer (Poly2) is shown in red, and the gold layer (Metal) is shown in black.  Because many of our structures are free-floating, the second polysilicon layer (Poly1) is hard to see on this drawing.  Probe points are labeled in Poly0.

Figure \ref{fig:device-drawing} shows the design: a simple, transverse, pawl and shuttle system, provides the ratcheting system for the grating.  By synchronized motions of opposing sets of actuator banks, the pawls drive the shuttles.  The design is axially symmetric, the two sets of opposing actuators surround the central shuttle system.  The shuttle is connected to a diffraction grating (not shown.)  The shuttle slides up and down on a track, constrained by small pill-boxes.

\begin{figure}[htp]
\begin{centering}
\includegraphics[width=5.5in]{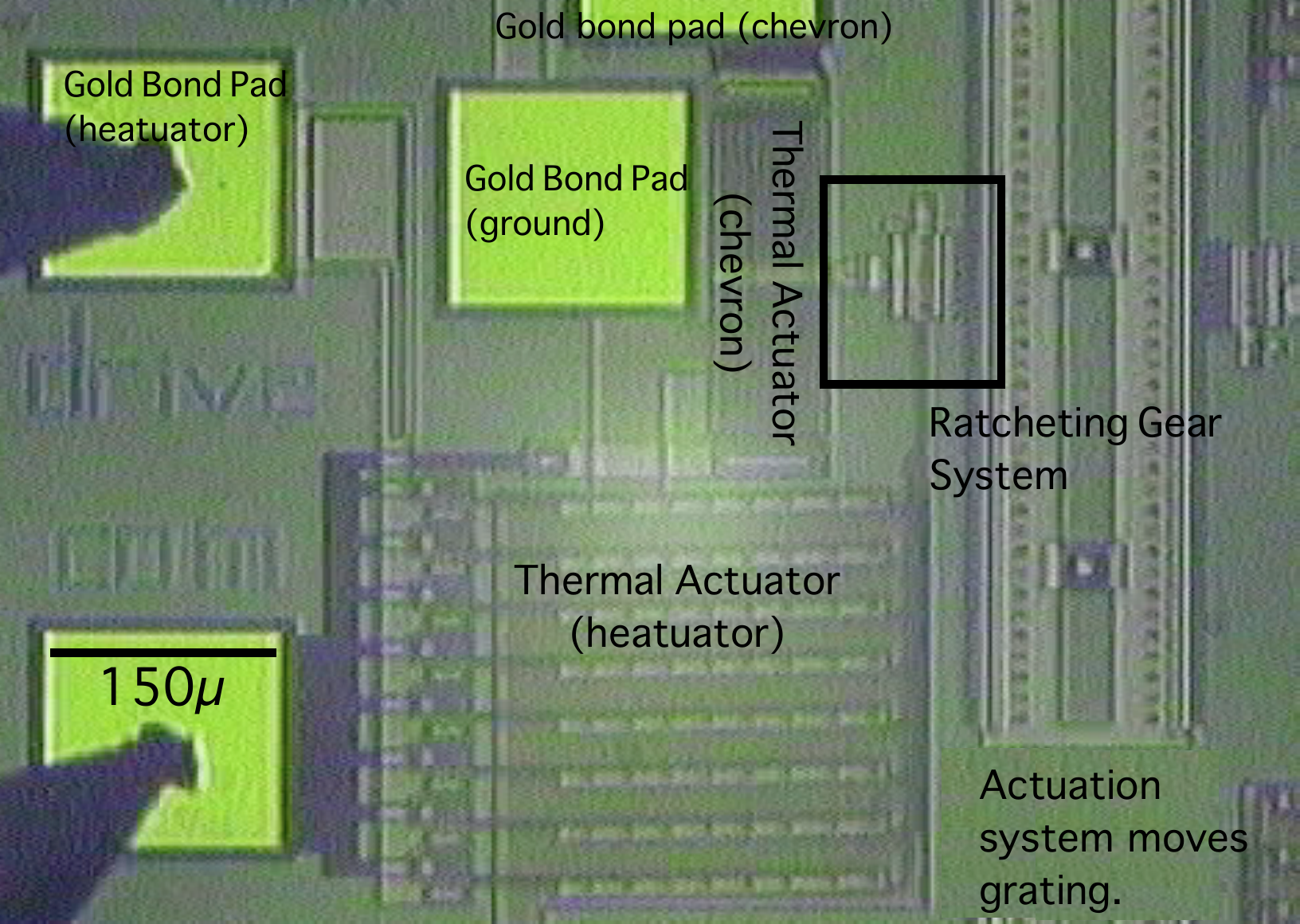}
\caption{\label{fig:zoom1}One half of the actuation system is shown on a prototype chip.  Two sets of thermal actuators are seen, the heatuator, and chevron.  In this picture, probes are sending current through the heatuator, causing thermal deformation.  This deformation presents as a slight tilt (from horizontal) in the heatuator bank.  Figure \ref{fig:pawlshuttle} zooms into the pall/shuttle system.}
\end{centering}
\end{figure}

\begin{figure}[htp]
\begin{centering}
\includegraphics[width=5.5in]{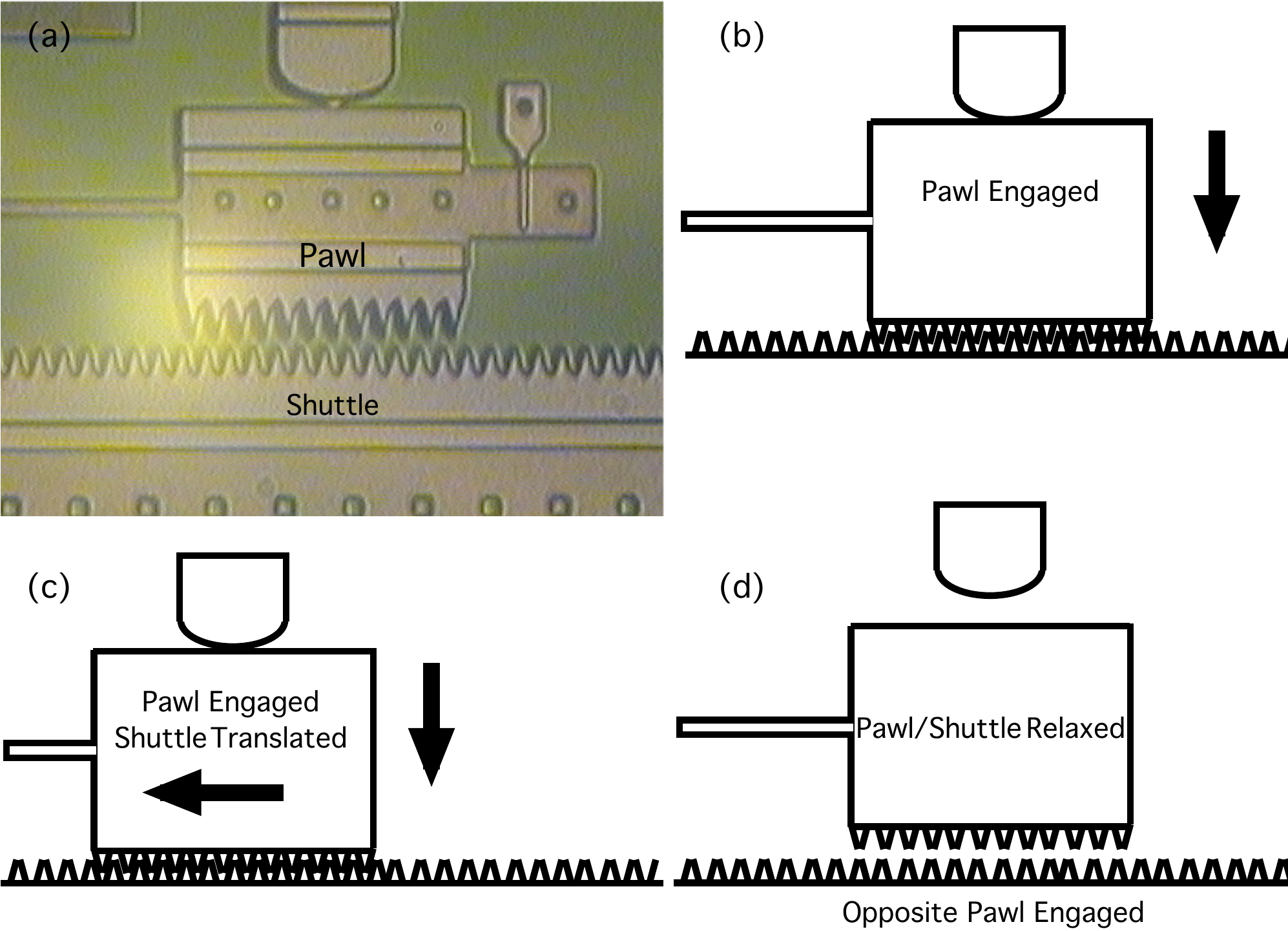}
\caption{\label{fig:pawlshuttle}(a) a close-up shot of the as-built pawl and shuttle system.  Some bridging of Poly2 is shown between the pawl and drive head.  This bridging does not appear on all chips. The cartoon diagram (b), (c) and (d) demonstrates the operation of the pawl/shuttle system.  (a) The system begins in a relaxed state, (b) the pawl is engaged by the chevron actuator (not shown.)  (c) The shuttle is translated by the heatuator bank. (d) The opposing side pawl engages, and this side's pawl/shuttle system relaxes.  (d) and (a) are equivalent.  These set of motions stretch and relax the diffraction grating.}
\end{centering}
\end{figure}

We now zoom into the design to explore the actuators.  Figure \ref{fig:pawlshuttle}, presents an as built system, with two sets of thermal actuators.  The bottom set are colloquially known as ``heatuators.''  Heatuators are U-shaped beams with a thick and thin side.  Because of differential heating on the thick and thin side, thermal deformation pulls the pawl.  The top thermal actuator is known as a ``Chevron Actuator.''  This actuator has parallel beams, bent in a V-shape, towards the pawl.  As current passes through the thermal actuator, the actuator pushes the pawl into the shuttle.

The actuator and pawl is actuated by a simple set of locking and pushing maneuvers to move the shuttle.  This set of maneuvers is illustrated in Figure \ref{fig:pawlshuttle}.  The sequence is:  
\begin{enumerate}
\item Lock the pawl into the shuttle. (b) Chevron engaged.
\item Translate the pawl.  (c) Chevron and heatuator engaged.
\item Engage opposite pawl.  (not shown) Chevron and heatuator engaged.  Opposite chevron engaged.
\item Disengage pawl from shuttle.  (d) Heatuator engaged.  Opposite chevron engaged.
\item Relax pawl. (a) Opposite chevron engaged.
\item Repeat step 1.
\end{enumerate}

After one set of these locking and pushing maneuvers is completed, the shuttle is translated by one set of teeth ($10\mu.$)  The grating is locked in place by engaging the pawl with the chevron actuator.

\subsection{The Built System}
The chips returned from PolyMUMPS were tested in July of 2007.  The pawl chevron actuator is driven by a square wave, under power-amplification, at 20$V$ using a variety of current-limiting resistors (logarithmically spaced from $1M\Omega$ to $100\Omega$.)  Unfortunately, the chevron actuator is too stiff and can not fully drive the pawl to be engaged with the shuttle.

The heatuator actuator is driven similarly. At $20V$ with the same set of current-limiting resistors.    When driven by the heatuator, the pawl travels further than a single gear-tooth ($\sim11\mu.$)  It is likely that the heatuator can drive the shuttle.  A number of test structures to measure heatuator performance independent of the chevron are on our chip, however, the chevron actuator is too stiff to test in the completed system.

The diffraction grating is fully compliant.  Tests on the grating via the probe indicate exceptionally good performance (within measurement uncertainty of several $\mu.$)  Our concerns were with the constancy of the grating spring constant (see \S\ref{S:grating}) and the repeatability of grating fabrication.

The diffraction grating has long runs of think polysilicon ($70\mu m\times3\mu m \times 2\mu m$.)   These long runs are sensitive to misalignment in the lithographic process.  Of the twenty diffraction gratings on the ten released chip, only one grating arrived broken, and may have been damaged during handling and test.

\section{DESIGN RISKS AND FUTURE DIRECTIONS} \label{S:risk}
As described in \S\ref{S:grating}, FEA is performed in equilibrium.  Temporal FEA has not been performed.  Typically, because MEMS devices are small, resonant frequencies are in the kilo- or mega- Hertz regime.  However, compared to typical MEMS, the designed grating is an enormous spring.  Temporal FEA should be performed, and its results fed into our grating efficiency simulation.

Because of the internal crystal stresses associated with etching polysilicon layers and the bulk micromachining process, the resulting diffraction grating may not be of ``optical quality'' ($\lambda/8$ surface quality.)  In addition, bulk micromachining limits the grating blaze angle to 90$^\circ$.   Both flaws can be mitigated by starting with a well-polished wafer, and using an etch based manufacturing process\cite{mar06,tormen06}.  As a result of silicon's crystal structure, and the readily available supply of wafers with crystal faces orthogonal to the wafer's plane, the blaze angle is fixed at 54.7$^\circ$.

Because of the chemical vapor deposition (CVD) process used to deposit polysilicon in the PolyMUMPS process, the resulting diffraction grating may not have surfaces that are of optical quality ($\lambda/8$.)  The Sandia Ultra-planar, Multi-level MEMS Technology 5 (SUMMiT V) Fabrication Process is a five-layer polycrystalline silicon surface micromaching process that uses chemical mechanical polishing (CMP) to obtain optical quality surfaces.  SUMMiT V is a likely candidate for the next generation fabrication run of the tunable grating.

PolyMUMPS design rules require that the pawl and shuttle gears are not fabricated in a meshed configuration.  As a result, the translator is locked only when a thermal actuator is pushing the gears together.  Thermal photons from the engaged actuator would add background noise to our NIR detector.  The SUMMiT V process allows gears to be fabricated meshed, and so this thermal background will disappear.

Packaging and assembly of a full optical design, though not an impossible task (see Axsun technologies for a design with similar level of complication) has not been considered for this design.  Future work will require careful consideration of packaging and test.

The resolution of a spectrograph is $R=m\cdot N$ where R is the resolution, m the order and N the number of illuminated grating elements.  The system requirements in \S2, indicate that we need R=4000.  By pushing up $m$, it is possible to significantly increase resolution while reducing the grating size, however, the trade-off is that orders begin to overlap, and some kind of order-blocking filter is necessary.  Therefor, to reduce the complexity of the design, small values of $m$ are valued.  We have not fully explored the value of $m$ for the grating.

MEMS design experience dictates\footnote{For example, see the web page: \url{http://www.memsrus.com/documents/PolyMUMPsNotes.Microsoft.pdf} } the designs which rely on friction, gears, or stiction, often have low yields and are not robust to repeated use.  Our design, due to the conflicting requirements of large displacement and large force, is a gear-based design.  A fully compliant\cite{kota01} MEMS design, cleverly using geometric advantage, may be more robust.

In summary, we present a new, MEMS, tunable diffraction grating.  This grating sits at the heart of a miniature, fiber-fed, NIR, medium-resolution, spectrograph.  The advantage of miniaturizing spectrographs comes from economies of scale.  By reproducing the same spectrograph $10,000\times$, the total cost of an instrument is drastically reduced.

\acknowledgments
 
We acknowledge the UCSC MEMS laboratory.  In addition, the authors acknowledge Mahyar X Fotovatjah, Stefan Gadomski, and Chung Quang Thai.

\bibliography{article}   
\bibliographystyle{spiebib}   

\end{document}